\documentclass[preprint,authoryear]{elsarticle}
\usepackage[T1]{fontenc}
\usepackage{ae,aecompl}
\usepackage{graphicx}
\usepackage{footnote}
\usepackage[fleqn]{amsmath}
\usepackage[varg]{txfonts}
\usepackage{multirow}
\usepackage{subfig}

\journal{Astronomy and Computing}

\begin{document}

\begin{frontmatter}

\title{PySE: Software for Extracting  Sources from Radio Images}

\author[api,ttu]{D.~Carbone}
\ead{dario.carbone@ttu.edu}
\author[cfa]{H.~Garsden}
\author[cwi]{H.~Spreeuw}
\author[uow]{J.~D.~Swinbank}
\author[gwu]{A.~J.~ van~der~Horst}
\author[api,astron]{A.~Rowlinson}
\author[astron]{J.~W.~Broderick}
\author[moca,moca2]{E.~Rol}
\author[berkley]{C.~Law}
\author[api]{G.~Molenaar}
\author[api]{R.~A.~M.~J.~Wijers}

\address[api]{Anton Pannekoek Institute for Astronomy, University of Amsterdam, Postbus 94249, 1090 GE Amsterdam, The Netherlands}
\address[ttu]{Department of Physics and Astronomy, Texas Tech University, Box 1051, Lubbock, TX 79409-1051, USA}
\address[cfa]{Harvard-Smithsonian Center for Astrophysics 60 Garden Street, Cambridge, MA 02138, USA}
\address[cwi]{Netherlands eScience Center, Amsterdam, The Netherlands}
\address[uow]{Department of Astronomy, University of Washington, Box 351580, Seattle, WA 98195-1580, USA}
\address[gwu]{Department of Physics, The George Washington University, 725 21$^{st}$ Street NW, Washington, DC 20052, USA}
\address[astron]{ASTRON, The Netherlands Institute for Radio Astronomy, Postbus 2, 7990 AA Dwingeloo, The Netherlands}
\address[moca]{Monash Centre for Astrophysics (MoCA), Monash University, Melbourne, Victoria, 3800, Australia}
\address[moca2]{School of Physics and Astronomy, Monash University, Melbourne, Victoria, 3800, Australia}
\address[berkley]{Department of Astronomy and Radio Astronomy Lab, University of California, Berkeley, CA, USA}

\begin{abstract}
PySE is a Python software package for finding and measuring sources in radio telescope images.
The software was designed to detect sources in the LOFAR telescope images, but can be used with
images from other radio telescopes as well.
We introduce the LOFAR Telescope, the context within which PySE was developed,
the design of PySE, and describe how it is used.
Detailed experiments on the validation and testing of PySE are
then presented, along with results of performance testing. We discuss some of the current issues with the
algorithms implemented in PySE and
their interaction with LOFAR images, concluding with the current status of PySE and its future development.
\end{abstract}


\end{frontmatter}

\section{Introduction}
\label{sec:intro}
The LOFAR Radio Telescope \citep[][]{LOFAR_vanHaarlem2013} is a radio interferometer comprised of many
antennae situated throughout Europe, and linked by a high-speed network. It is one of the new generation radio
telescopes, along with 
the Australian Square Kilometer Array Pathfinder \citep[ASKAP;][]{ASKAP_Johnston2008}, the Murchison
Widefield Array \citep[MWA;][]{MWA_tingay2013}, and the Long Wavelength Array \citep[LWA;][]{LWA_Ellingson2009}.
These telescopes provide high-resolution wide-field imaging, and are proving grounds for the Square
Kilometre Array (SKA),
which will become the most sensitive radio telescope ever built.
LOFAR observes at the low end of the radio spectrum, between 30 and 250\,MHz, and its capabilities are being used
to support six important projects in radio astronomy, of which transients and variables are one.
In order to discover transients, the LOFAR Transients Key Science Project
\citep[TKP; ][]{LOFAR_Fender2008} has developed a Transient Pipeline \citep[{\sc TraP};][]{Trap_Swinbank2015}
that is able to search for transient and variable sources
in an image stream in real time. One of the main steps of this pipeline is the extraction of all the sources present in
each image as quickly and as accurately as possible, with a main focus on point sources.
Transient sources would very likely be point sources, because a transient typically has a small angular
size on the sky, much smaller than the resolution of the instrument.
It is within this project that the PySE (Python Source Extractor) has been developed.

Transient signals observed by radio telescopes are generally classified as fast or slow
\citep[][]{Cordes2004, LOFAR_Fender2008},
with the boundary between the two classes set at  $\sim$1 second, and fast transients occurring down to microseconds.
Fast radio transients may be produced by pulsars, gamma ray bursts, flaring stars, and fast radio bursts
\citep[][]{FRBs_Lorimer2007}. Slow transients are produced by large scale high energy events such as flares in
quasars, tidal disruption events, and supernovae.
There are likely to be unknown transient types in both classes yet to be discovered.

Fast transients are usually too short to appear in an image as their flux would be averaged to very low values; therefore
dedicated high speed hardware and software \citep[e.g.][]{Seryalk:2013} have been developed to detect them.
Slow transients, that appear for long enough to be detectable in a image, can be detected by scanning an image;
they will appear as an unknown astronomical object when checked against catalogues
\citep[e.g.][]{Adam_Stewart2015, GCRT1745_Hyman2005}.
This process is therefore one of finding sources in an image, determining their location,
brightness, and other parameters, and matching them with known sources. These tasks are well-handled using
computer software that detects and characterises sources, and databases that
do the data comparison. PySE is an instance of a source finder
designed especially for detecting slow transient (i.e. point sources) in a series of images.

This paper describes the 
functionality of the PySE source finder, presents experimental results indicating how
accurate it is, and some performance measures. 
Section~\ref{sec:reason} illustrates the reasons for creating a new source finder algorithm,
Section~\ref{sec:the_program} describes the program,
Section~\ref{sec:simulations} presents different methods to generate simulated maps used to test PySE,
Section~\ref{sec:sftests} reports on experiments that demonstrate the accuracy of PySE, Section~\ref{sec:performance_results}
indicates how PySE performs, and Section~\ref{sec:discussion} discusses our intentions for further use and development.

\section{Need for a new source finder}
\label{sec:reason}

PySE was designed in order to operate as a source finder for the LOFAR Transients Key Project
and to be integrated within the {\sc TraP}\footnote{For version 3.0, see http://tkp.readthedocs.org/en/r3.0/index.html}.

Investigations into the acquisition or development of a source finder for the Transients Key Project have been performed 
and some requirements have been set such as:

\begin{itemize}
\item Speed: the source finder should run within few seconds per image.
The run-time will depend on the size of the image and on the number of sources, with typical images being around
$1024^2$\,pixels containing about 100 sources
corresponding to a run-time of $\lesssim$\,3\,seconds
\item Accuracy: the source finder should calculate the background noise and the source parameters
as near as possible to theoretical accuracy limits
(background noise within 1 percent, positions within 1 pixel, and fluxes within 5 percent, in absence of noise and within
20 percent in presence of noise and other effects).
The source parameters should also have robust error estimates.
\item It should be able to handle the correlated noise generated by radio telescopes \citep[][]{Wilson:2012}.
\item It should be robust, since it will be run as part of an automated imaging and transient-detection pipeline;
it must be able to handle images of varying quality, deal with problems, and fail gracefully when failure is absolutely necessary.
\end{itemize}

Based on these requirements, existing source finders were evaluated as candidates for a LOFAR transients source finder.

\subsection{Source finder evaluation, and the impetus for a new source finder}

Several programs have been developed for extracting sources from an astronomical image, in most of the cases to
fulfill specific requirements of the developers team.
Source finders optimised to work on optical or infrared images, e.g. SExtractor \citep[][]{Bertin:1996} or 
CuTEx \citep[Curvature Threshold Extractor;][]{cutex_Molinari2011}
are not able to deal with correlated background noise that is typical of radio images
and were therefore excluded from the candidates for a LOFAR transients source finder because they could not meet
our requirements.

Other source finder tools have recently been developed to extract as precisely as possible all the emission present in
a radio image, for example
PyBDSF \citep[Python Blob Detection and Source Finder, formerly known as PyBDSM;][]{Mohan2015},
Duchamp \citep[][]{duchamp_Whiting2012}, SOURCE\_FIND \citep[][]{source_find_Franzen2011},
and Aegean \citep[][]{Hancock:2012}.
These algorithms, which were developed in parallel with PySE, do not focus on the speed of processing large datasets.
They proved to be too slow and did not match our speed requirement.

For the above mentioned reasons, it was decided to develop a new algorithm, suited for the purposes of the LOFAR
Transients Key Project: PySE.

\section{The PySE Program}
\label{sec:the_program}

\subsection{PySE design}
\label{sec:pyse_design}

PySE has been developed to satisfy the requirements described in Section~\ref{sec:reason}.
The source-finding algorithms are contained in a stand-alone software package that can also be integrated into 
a pipeline, specifically the LOFAR {\sc TraP}.
PySE is designed to process typical LOFAR images (1024$\times$1024 pixels containing about 100 sources) in
3 seconds \citep[][]{Spreeuw:2010}
\footnote{The software described in \citep[][]{Spreeuw:2010} discusses a version prior to 3.0, that is identical to what
described in this work, apart from the addition of the features of forcing the shape of sources to be the same as the
synthesized beam (see Section~\ref{sec:run_pyse} ) and to extract the flux from predetermined coordinates (see
\ref{app:option}).}.
For the processing of images in the {\sc TraP}, a real-time, distributed,
parallel task scheduling system has been installed, which facilitates the running of many instances of the source
finder concurrently, providing the necessary throughput from an automated source-detection pipeline.
PySE is built in a high-level programming language, Python, with a modular design,
using software engineering practices such as issue-tracking, source control, unit testing, and has been
extensively tested (this paper; \citealt{transsel_Rowlinson2016, Trap_Swinbank2015}). It has built-in
exception-handling and failure modes.

The source-finding steps are as follows:
\begin{enumerate}
\item Find the RMS noise and background over the image
and subtract the background map from the original image.
\item Find islands of high pixel values above the noise. From these, create a list of sources for analysis.
\item Deblend islands that may contain multiple sources.
\item Fit parameters to sources. The source parameters are based on 2-D Gaussian fitting to the source.
\end{enumerate}

In order to estimate the background and the noise of an image (Step 1), it is divided into boxes. In each box {\em k},
$\sigma$ clipping around the median is performed until it has converged, i.e., no further pixels are excluded.
{\em k}, $\sigma$  clipping has been implemented in PySE slightly differently from SExtractor, by clipping
$\pm$\,{\em k}\,$\sigma$ around the median instead of $\pm$\,3\,$\sigma$. The value of {\em k} is calculated
at each iteration and depends on the number of independent pixels in the image at every step (see e.g., Section~7.1
of the SExtractor manual).
At this point the mode and the standard deviation of the distribution of pixels are calculated. The mode is calculated
using Equation~\ref{eq:mode} if the distribution of pixel values is not too skewed otherwise the mode is assumed to
be equal to the median, as SExtractor does.

\begin{equation}
mode = 2.5 \times median - 1.5 \times mean .
\label{eq:mode}
\end{equation}

The boundary between these two scenarios is set by the condition in Equation~\ref{eq:skew}.

\begin{equation}
\frac{|mean\,-\,median|}{sigma} \le 0.3 .
\label{eq:skew}
\end{equation}

The calculated modes and standard deviations in each box are than interpolated to produce the background and noise
maps respectively. The background map is then subtracted from the original image before sources are identified and extracted.

In order to identify sources (Step 2), the noise map is multiplied by a user specified number (the so-called {\em detection}
threshold). Pixels with a value higher than the local threshold level are considered as source pixels. Once sources have
been identified, neighboring pixels with value above another, lower, threshold (determined multiplying the noise map by
a user specified number, the so-called {\em analysis} threshold) are associated to the sources.
In case the False Detection Rate algorithm \citep[FDR;][]{Hopkins:2003} is used, the analysis threshold is not calculated
and is set equal to the detection threshold. The analysis threshold can never be higher than the detection threshold.
The FDR algorithm is a method that allows to impose a maximum fraction of pixels being falsely identified as sources.
It calculates internally the best value of the detection threshold to be used in order to fulfill the required limit. 
The implemented FDR algorithm divides the initial map (after subtracting background) by the noise map. This normalized
map is then used to calculate the detection threshold, as explained in Appendix B of \citet{Miller:2001}.
After this, the source extraction works as explained above.

If two or more sources are close enough, they will belong to the same island and will initially be detected as one source.
The deblending algorithm (Step 3) implemented in PySE is analogous to the one used by SExtractor. The user specifies
the number
of subthresholds exponentially spaced between the lowest and the highest pixel value in each island. If the two (or more)
sources are separated by any of these subthresholds, they will be recognized as separate sources if in each of them there
is at least one pixel with value above the detection threshold.

2-D Gaussians are fitted to the detected sources in order to estimate their positions (right ascension, declination), their shapes
(semi-major axis, semi-minor axis and the position angle of the semi-major axis, east from local north) and their brightness
(peak flux density), with respective errors. The integrated flux density is reported as well, even if it can be derived from the
peak flux density, the size of the axes and the size of the synthesized beam. The fit of the source parameters is performed
using a least-squares convergence algorithm, in particular a modified version
of the Levenberg-Marquardt algorithm \citep[][]{Levenberg:1977}. Correlated noise, and errors derived with Gaussian fits
in the presence of correlated noise, are implemented from theory \citep[][]{Bertin:1996, Condon:1997, Refregier:1998,
Fomalont:1999, Spreeuw:2010}.
The user may also request for a residual map, where, for each island, the computed Gaussian has been subtracted from the data.

\subsection{PySE Execution}
\label{sec:run_pyse}
PySE is a standalone Python program that uses the source-finding modules. It is run with several options, followed
by a list of input image files, which can be in standard FITS or CASA\footnote{www.nrao.com} format. On
termination the program prints a list of found sources, and may be asked to save information such as the
calculated background  image, the source list in different formats (e.g. text, region files importable in
DS9\footnote{ds9.si.edu}), and other data requested via command line options.

The main options of PySE are explained in this Section, while a detailed description of all the options is given in \ref{app:option}.

The detection threshold is defined as a multiple of the local noise. Pixels lying above this threshold are defined as sources.
Neighbouring pixels brighter than a secondary, lower threshold, the analysis threshold, are grouped into islands and fit
to a 2-D Gaussian to extract the source parameters.

The grid size is a very important parameter. It is a value, in number of pixels, that is used in the estimation
of the background and the variance map.
In order to calculate the noise, the image is split into squared cells of size equal to the grid size.
The rms in each cell is calculated and these values are interpolated to create the background mean and variance map.
Changing the grid size will therefore potentially affect the noise estimation and the source extraction.
The grid size should not be smaller than the size of individual sources, otherwise their flux will dominate the rms
calculation in individual cells and be mistaken as noise and 
both the background mean and variance will be overestimated.
It should not be too coarse as well, otherwise the interpolation will not adequately follow the noise pattern in the image.
In most cases, i.e. when sources are generally fairly compact or only moderately resolved, 
a grid size of a few times the most extended source would work adequately.

Another very important option of PySE is the forced clean beam fit.
Using this option the shape (major and minor axes and position angle)
of the extracted sources is not fit, but forced to be the same as the restoring beam. This implies that all of the extracted sources
will have the shape of point sources. This option is especially important in the {\sc TraP}. In an image stream a source
might appear suddenly much brighter because an image is noisier, and the source finder might misfit it with a much larger
2-D Gaussian introducing artificial variability in the light curve we create in {\sc TraP}, making it more difficult to identify real
variables. Forcing the shape of the source would fix such issues.

The {\sc TraP} will build light curves for all the extracted sources
through positional matching.
If a source is not detected in an image, either
because it is variable or because an image is noisier, the pipeline will extract the flux of that source at its position anyway.
This is done using the forced beam shape with fixed position fit
option in PySE. This option allows the extraction of the flux at specified coordinates
even if there are no pixel values above the detection threshold at that position.

\section{Simulations}
\label{sec:simulations}
We used different methods to produce simulated maps to test our software, which are
targeted to test particular aspects of the program in detail.

\subsection{Method 1: noise and point sources only}
\label{sec:Hanno}
First, we want to test the software accuracy in 
measuring background noise and in
recovering sources fluxes and positions. To do so, we have created three types of simulated maps, starting in all cases
from the same, real LOFAR observation.
All resulting images have 1024$\times$1024 pixels and resolution of about 2.5\,arcmin (5 pixels).

The first type is constituted of pure correlated noise, the second contains point sources on a zero background,
while the third has point sources on a non-zero background.

In the first case, we empty the image
(in the UV plane, therefore preserving the sampling information from the original real observation),
generate Gaussian noise
with rms 1\,Jy
in the visibilities and image them by applying a Fourier transform.
This way the noise in the image plane is correlated.
We produced 10$^4$  individual maps that will be used to test the accuracy of PySE in estimating the
background noise in a radio image.

To produce the second types of maps, we empty the image
as in the previous case
and insert 100 point sources
with flux between 2 and 20\,Jy, 
with the size and shape of the appropriate restoring beam directly in the images.
The sources were inserted at random positions, provided they were more than 3 Half Width at Half Maximum (HWHM) apart.
These maps therefore have point sources on top of a zero background. They will be used to test the accuracy of PySE in
recovering sources fluxes and positions.

In the third case we produced correlated noise, as explained for the first type.
After that, 
100 point sources, with flux between 2 and 20\,Jy and size and shape of the appropriate restoring beam, were added
into the images. These maps will used to test the ability of PySE to retrieve information on sources in a real case, with
sources on top of correlated noise.

We have combined the results from the source extraction in each separate image to have statistics on the results in
order to avoid having blended sources.
We have used these images to validate the code in Section~\ref{sec:validate_pyse}.

\subsection{Method 2: simulating a large number of sources}
\label{sec:Hugh}
We have created an entirely different set of simulated images with the only purpose of testing the speed of the algorithm.
We created a pixel array containing point sources and random noise, which is then convolved with a clean beam
(a 2-D Gaussian) and saved as an image.
The images are square, with widths in pixels between 128 and 10240, containing 4 up to 25600 sources.
Point sources are placed on a regular grid
within the image, so they are equally spaced from each other. The smallest image is of width 128 pixels, containing
4 sources; subsequent images are larger, and have the same density of sources.
The grid separation ensures that sources are not blended.
While the arrangement of the sources is not astronomical, it is irrelevant for the source finder operations. The fluxes
of the point sources are all the same. Since we are only interested in PySE speed, the images have been
designed to make the sources all identical (same shape, size and flux) and easy to find (sources are not blended and
much brighter than the noise).
The full list of images size and number of sources is given in Table~\ref{tab:widths}.
These images were used to test the speed of PySE in Section~\ref{sec:performance_results}.

\begin{table}
\centering
\begin{tabular}{|r|r|r|}
\hline
Image Width (pixels)		& Number of Sources &  PySE Run Time (sec) \\
\hline
128					& 4		& 1.08 \\
256					& 16		& 1.37 \\
512					& 64		& 2.57 \\
1024					& 256	& 7.68 \\
1536					& 576	& 18.74 \\
2048					& 1024	& 26.91 \\
3072					& 2304	& 58.85 \\
4096					& 4096	& 106.59 \\
5120					& 6400	& 164.12 \\
6144					& 9216	& 239.47 \\
7296					& 12996	& 331.67 \\
8192					& 16384	& 433.35 \\
9216					& 20736	& 525.73 \\
10240				& 25600	& 624.89 \\
\hline
\end{tabular}
\caption{The width of the images used for PySE speed testing, and the
number of sources in each image. The number of sources is proportional to 
the width squared and hence to the total number of pixels.}
\label{tab:widths}
\end{table}

\subsection{Method 3: simulating LOFAR maps}
\label{sec:Dario}
Finally, the last set of simulated images is created to simulate real observations,
including, for example, calibration and imaging errors.
These simulated images were generated to have several
characteristics of real LOFAR data, using the following procedure. We start with an existing LOFAR observation in
which we replaced the visibilities with Gaussian noise
as in Section~\ref{sec:Hanno}
The standard deviation of the background noise
was created to replicate one of a real observation, and as such it
depends on the observing frequency,
bandwidth, integration time, and configuration of the  array when the observation was taken.
The noise per baseline (N) is given by:

\begin{equation}
\centering
N_{i, j} = \sqrt{\frac{\text{SEFD}_i \times \text{SEFD}_j}{2 \, W \, I}} \: ,
\label{eq:noise}
\end{equation}

\noindent with

\begin{equation}
\centering
\text{SEFD} = \frac{2 \, \eta \, k_\mathrm{B}}{A_\mathrm{eff}} T_\mathrm{sys} \: ,
\label{eq:SEFD}
\end{equation}

\noindent where $SEFD_i$ and $SEFD_j$ are the System Equivalent Flux Densities for each station,
$W$ is the total bandwidth,
$I$ is the integration time, $k_\mathrm{B}$ is the Boltzmann's constant, and $\eta$ the system efficiency factor
($\sim$1.0). $A_\mathrm{eff}$, the effective collecting area, and $T_\mathrm{sys}$, the system noise temperature,
are dependent on the array configuration and the observing frequency.
In our case, the standard deviation of the background noise is about 1\,Jy.
We have created our own set of sources to be input in the simulations.
This included 100 point sources per image, with fluxes between 2 and 20\,Jy.
The flux distribution of the sources follow a power-law distribution with exponent equal to -1.5.
We added the simulated sources into our dataset using BlackBoard Selfcal \citep[BBS;][]{vdTol:2007,Loose:2008}.
This predicts how the sky model (the list of sources to be inserted) would have been observed by LOFAR given the
telescopes gains and frequency used for the original observation.

The simulated data are stored in a Measurement Set (MS);
each observation has its own MS, containing all the information pertaining to an observation.
An image is generated by passing the MS through a pipeline consisting of three main steps:
flagging, calibration, and deconvolution. Flagging removes bad data from the MS, such as radio-frequency interference
and bad data produced by antenna, network, and other failures. Flagging is not necessary for simulated data since bad
data are not currently simulated.
Calibration adjusts the flux levels of the data so that known sources are matched as closely
as possible, thus producing accurate fluxes throughout the rest of the field. 
In our case, it accounts for noise-induced inaccuracies
and allows the simulation to reproduce possible real inaccuracies introduced in this step.
Deconvolution \citep[][]{Hogbom:1974} is
necessary because the image is blurred by a point-spread-function (PSF) resulting from the small, limited number of
antennas that sample the signal; deconvolution removes the side lobes of the PSF and generates the final image. It should
be noted that the accuracy of source-finding rests on the accuracy of the aforementioned steps, as source-finding operates
on the image produced by them. In our experiments, we assume that these steps have themselves been validated and their
accuracy proven. 
Deconvolution and image generation is performed by
AWImager \citep[][]{awimager_Tasse2013}.
The images created using this method are 512$\times$512 pixels, contain 100 sources each
and are used to estimate the best values of the parameters to be used in the {\sc TraP} in Section~\ref{sec:pyse_params}.

\section{The Accuracy of PySE: Validation and Testing}
\label{sec:sftests}
Once images have been made, the next step in transient searches is finding and characterising sources by using a
source extraction tool. Two main different steps have been taken to test the source finder performance: the first one is
validating the functionality of the source finder itself, while the second consists of the optimization of the parameters used
in the source extraction, especially when performed within the {\sc TraP}.

All tests performed in this document were performed with the version of PySE incorporated in version 3.0 of TraP,
released on 2015-12-14. The software is open source and is publicly available on GitHub as part of the TKP
software packages\footnote{https://github.com/transientskp/tkp/releases/tag/r3.0rc}.

\subsection{Validation of PySE}
\label{sec:validate_pyse}

We use the simulations described in Section~\ref{sec:simulations} for the validation of PySE.
In this section we want to test whether PySE is able to correctly measure the background and the noise in an image.
We also test the accuracy of the flux measurements performed by PySE, both in absence and in presence of a background
in order to distinguish where the flux uncertainties come from.
Finally, we estimate the completeness and reliability of PySE. These parameters have also been calculated in the
ASKAP/EMU Source Finding Data Challenge \citep{sfchallenge_Hopkins2014}, the results of which are summarized in
Section~\ref{sec:discussion}. The tests in \citep{sfchallenge_Hopkins2014} cover the performance of PySE on more
realistic radio maps, involving crowded fields and presence of extended sources.

\begin{figure}
\centering
\includegraphics[scale=0.5,viewport = 0 0 550 410, clip]{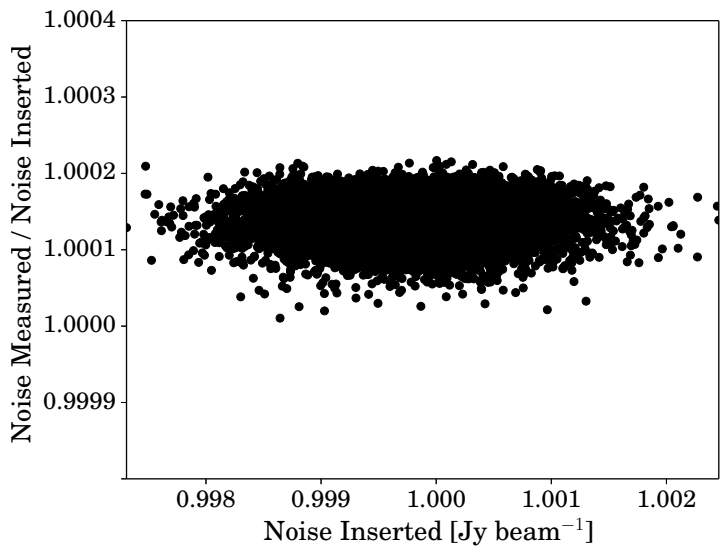}
\caption{
Validation plots of the RMS noise. The vertical axes represents the ratio of the measured and the inserted RMS noise.
The scatter around the value of one is less than to 0.1 percent.
}
\label{fig:noiseonly}
\end{figure}

We used the simulation procedure described in Section~\ref{sec:Hanno} for the validation of PySE.
In order to validate the software, we want to be sure that the noise measured by PySE is correct.
In the TKP pipeline, sources are detected above a threshold defined in terms of the local RMS noise. A robust estimation of the
local RMS noise is therefore necessary since any underestimate of the local RMS noise can cause false detections: that is,
noise peaks will be incorrectly identified as sources.
Therefore we ran tests on 10$^4$ images produced using maps of the first type described in Section~\ref{sec:Hanno},
in which the RMS of the noise is equal to 1\,Jy beam$^{-1}$.
The exact values of the inserted RMS are shown on the horizontal axis of Figure~\ref{fig:noiseonly} and are calculated as
the RMS of the distribution of pixel values in the maps.
We then ran PySE on these maps in order to verify that it is able to measure the correct noise from any image.
This is done by measuring the RMS of the pixel distribution of the background map generated by the algorithm.
The results of this test are summarised in Figure~\ref{fig:noiseonly} and show that the RMS noise we recover is very similar
to the RMS that was injected, with a deviation of less than 0.1 percent.
The measured noise is systematically overestimating the inserted one.
This can be explained by the fact that image pixels are not normally distributed, despite the fact that the visibilities were.
This is probably due to the fact not a proper Fourier Transform is applied, but a gridded Fourier Transform.

\begin{figure}[htbp]
\centering
\includegraphics[scale=0.31,viewport = 0 0 550 410, clip]{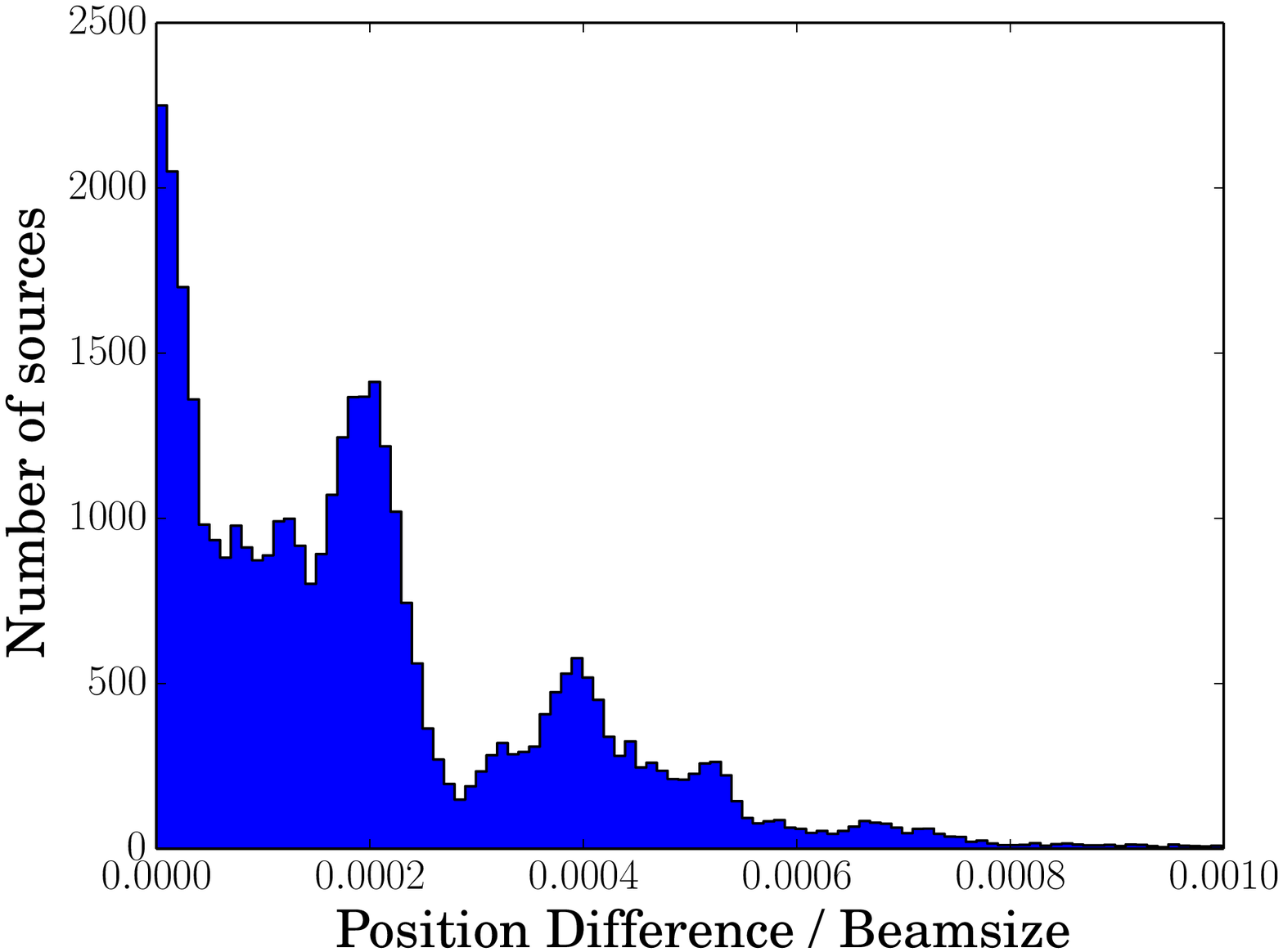}
\includegraphics[scale=0.31,viewport = 0 0 520 400, clip]{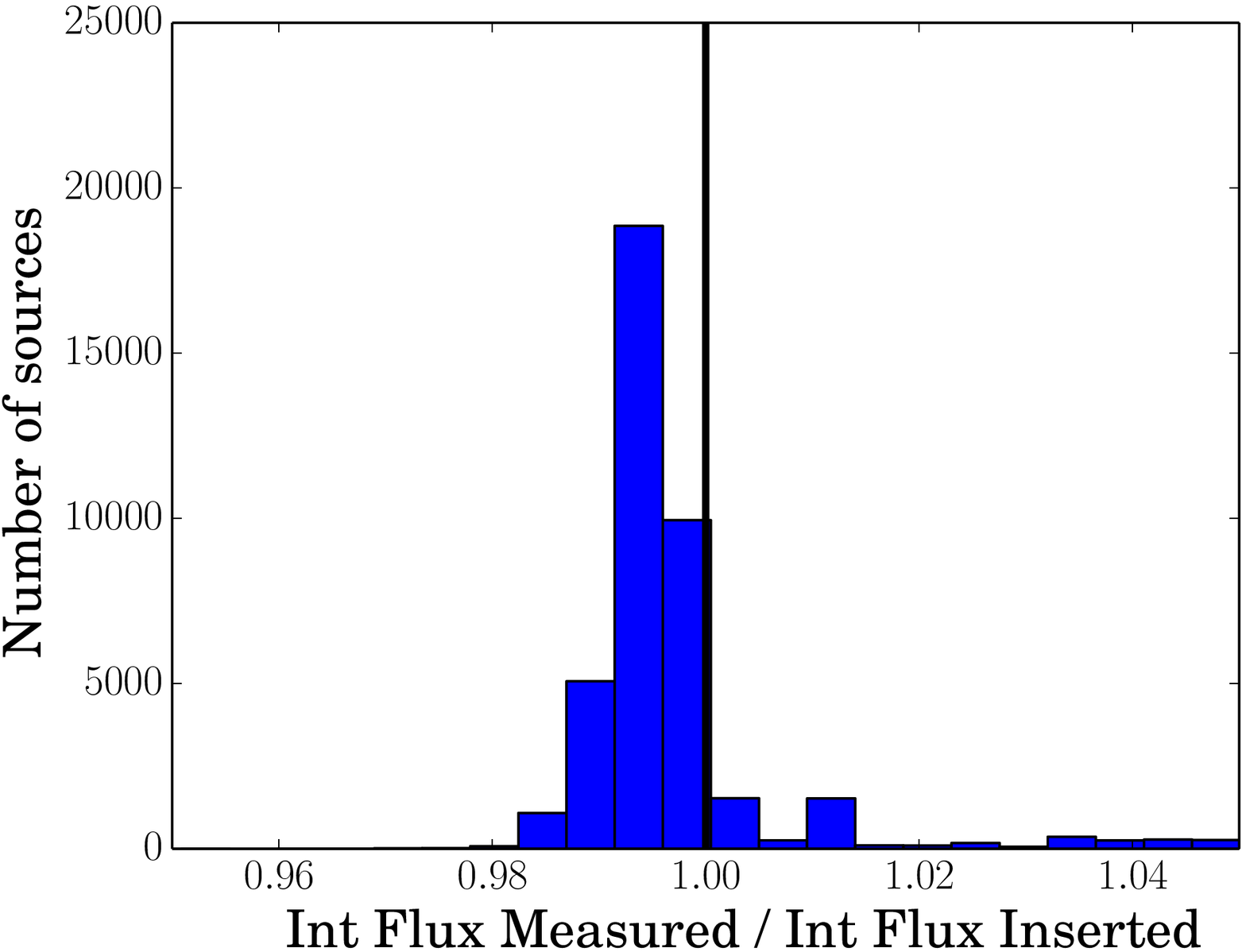}
\caption{
Validation plots of the positions (left panel) and integrated flux (right panel) on a total of 40000 simulated point sources
on a zero background. The vertical line in the right panel represents where the ratio between measured and inserted
integrated flux is equal to 1.
}
\label{fig:valid_flux_nonoise}
\end{figure}

The next step is to verify that PySE is able to extract sources at the right locations and with the right parameters (shape and flux).
To do this we first produced images with point sources on a zero background.
We inserted point sources with fluxes randomly distributed between 2 and 30\,Jy.
We have tested a total of 40000 sources.
The results show that PySE is able to reconstruct the right sources at the right locations and with the right parameters.
In Figure~\ref{fig:valid_flux_nonoise} we show the recovery of the positions of the extracted sources and their integrated flux.
The positions of the sources are measured with very good precision: the position difference is less than 0.1 percent of a beam.
The ratio between the measured and the inserted integrated flux is more spread around 1, even if the discrepancy is well
below 5 percent.
We note that there are more sources having an underestimated flux rather than overestimated.
This is due to the fact that the background mean (i.e. the average of the background noise) is generally overestimated
from {\em k}, $\sigma$ clipping due to the presence of sources which cause the background mean to be higher than
zero. This background mean is then subtracted leading the measured peak flux densities (and therefore the integrated flux)
to be too low.
The average source shapes (major and minor axes and position angle of the 2-D Gaussian) and the peak fluxes
are reported in the left half of Table~\ref{tab:ratios}.
We highlight that in this test we do not calculate errors on the extracted parameters because the error calculation
(both in fluxes and in positions) is based on the presence of a background noise and therefore not possible.
The values of the scatter reported in Table~\ref{tab:ratios} are fluctuations around the mean of all the measurements, not the
errors calculated by PySE. These fluctuations (as well as those in the fluxes) are due to
differences in the background interpolation: despite we used noise free maps, PySE still performs {\em k}, $\sigma$ clipping,
not eliminating all source pixels. These residuals are then interpolated to create a background map that is different for each
simulated image. 
We can see that all of the average values are very close to one.
The position angle is 
less well constrained as the average difference is 9 degrees with a large scatter;
this is caused by the fact that the beam in this dataset is almost circular
(major/minor axes of the beam = 1.04) and therefore it is harder to constrain its orientation.
We have calculated the completeness (fraction of the input sources being detected) and reliability (fraction of the detected
sources to be real) for our runs for each simulated image. The completeness is 99.99 percent: 2 sources
out of 40000 were not detected. The reliability is always 100 percent. 

\begin{figure}[htbp]
\centering
\includegraphics[scale=0.31,viewport = 0 0 550 410, clip]{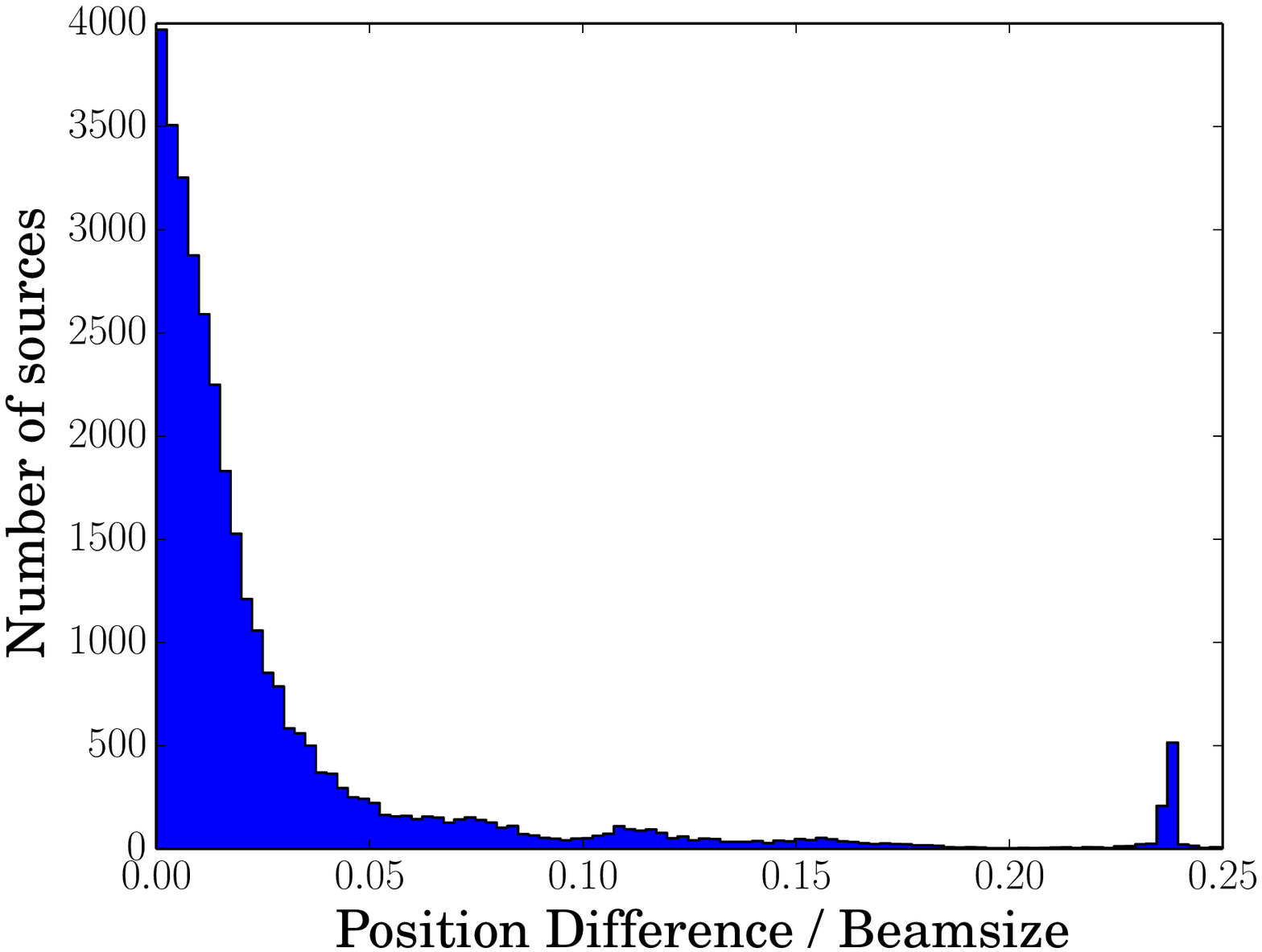}
\includegraphics[scale=0.31,viewport = 0 0 530 413, clip]{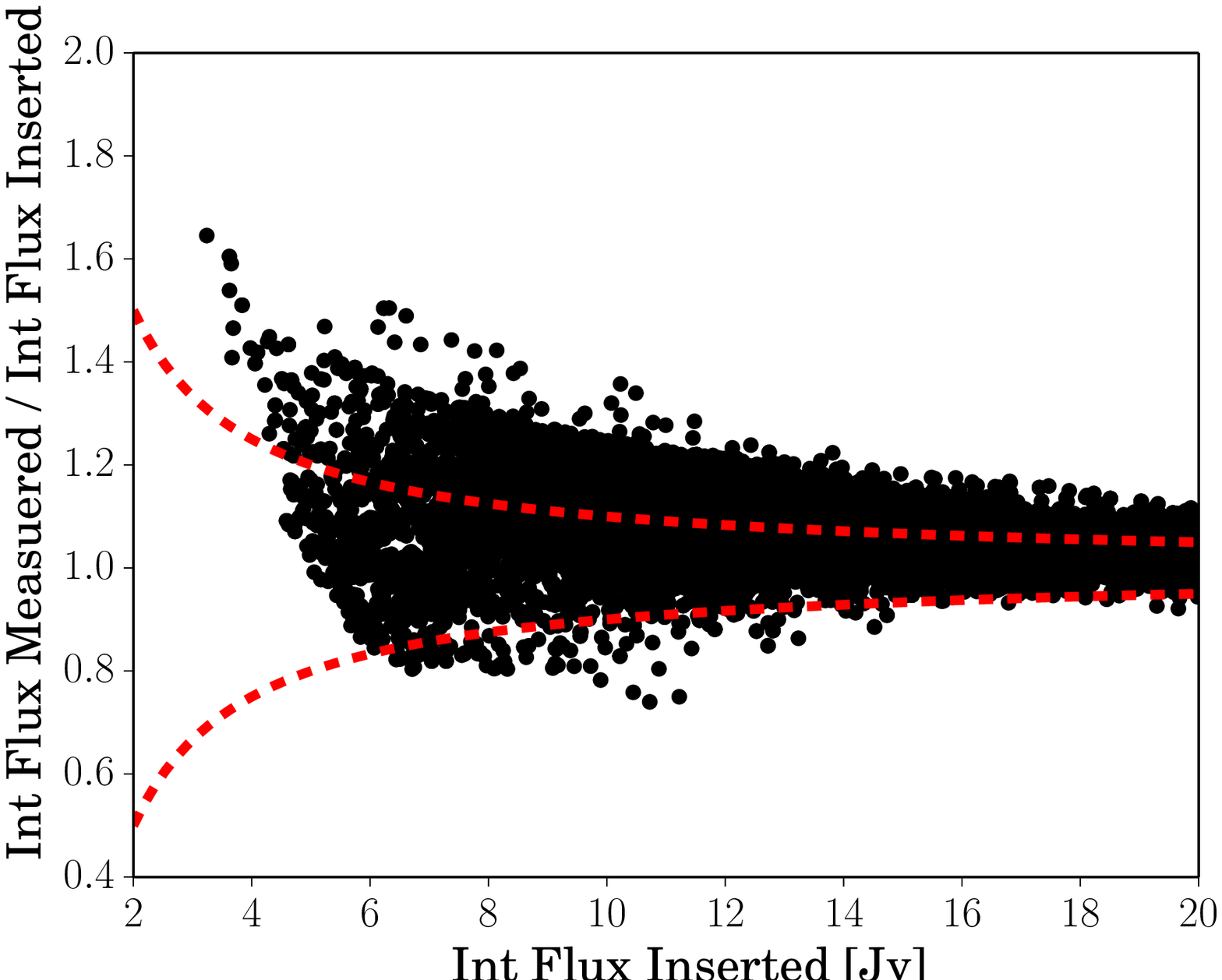}
\caption{
Validation plots of the positions
(left panel) and integrated flux (right panel) on a total of 40000 simulated point sources on a non-zero
background.
The dashed lines in the right panel represent the scatter around a ratio of 1 due to the noise level of 1\,Jy\,beam$^{-1}$.
}
\label{fig:valid_flux_noise}
\end{figure}

We then repeat this test with maps in which we added point sources to Gaussian noise.
\citet{Spreeuw:2010} showed that in the presence of sources, the background noise is overestimated.
This is due to the fact that, in the presence of sources, the lower tail of their distribution
will mix with the background noise in the image. The contribution of these faint sources will persist in the remaining pixels after
{\em k}, $\sigma$ clipping and, since sources are positive defined, they will bias the background mean, resulting in too much background
being subtracted. This effect was already pointed out by \citet{Condon:1997}, and PySE has implemented a correction to this effect
for the peak flux estimation.
In our test we added point sources on top of Gaussian noise with RMS equal to
1\,Jy beam$^{-1}$. We inserted point sources with fluxes randomly distributed between 2 and 30\,Jy
to simulate sources with both a low and a high signal-to-noise ratio.
We have set a detection threshold of 5 times the noise so we expect to extract sources brighter than about 5\,Jy.
The results of our test are shown in Figure~\ref{fig:valid_flux_noise} for positions and integrated fluxes, and in Table~\ref{tab:ratios}
for the average source shapes (semi major, semi minor axes and position angle of the 2-D Gaussian) and peak fluxes, as well as
their respective errors.
We observe that the scatter around the average value of the parameters reported in Table~\ref{tab:ratios} is higher than the one observed in 
the noise free case, as expected, but the averages are still close to one apart from the position angle that is not constrained.
In particular, we observe that the peak flux ratio is very close to one, despite the aforementioned underestimation of the background,
indicating that PySE is able to correct for this effect.
We also observe that even if it is still compatible with 1, the semi major axis ratio is on average overestimated.
This could be due to the presence of noise which causes the number of pixels associated to each source (pixels with values above
the analysis threshold, see Section~\ref{sec:pyse_design}) to increase, enlarging artificially the shape of the sources.
The same effect is not observed for the semi minor axis.
\citet{Condon:1997} mentioned the possible presence of biases in recovering Gaussian parameters, particularly at low signal-to-noise.
The average difference between the simulated and measured position angle is 66 degrees, with very large scatter.
The explanation of this discrepancy is again that the beam shape is almost circular, making more difficult to identify its orientation.
Observing the left panel of Figure~\ref{fig:valid_flux_noise} we can see that the positions are still very well determined.
There are a few outliers for which the position difference is around 0.25 beams, which correspond to one pixel.
The fluxes are scattered around the inserted values and the spread, measured as the standard deviation of the distribution,
is equal to the noise in the maps (1\,Jy).
The spread is due to the noise is indicated by the two dashed lines in Figure~\ref{fig:valid_flux_noise}.
All of the datapoints are consistent with the area delimited by the two lines within 1\,$\sigma$.
In the right panel of Figure~\ref{fig:valid_flux_noise} there seems to be an overestimation in the recovered integrated flux.
This is very likely due to the overestimation of the Gaussian parameters, especially the semi major axes, due to the presence of noise
as explained above. In fact, we have calculated the average value of the ratio between the measured and the inserted integrated flux,
and compared this number to the result of the product of the average ratios of the peak flux, the semi major and semi minor axes, as in
Table~\ref{tab:ratios}. They are both equal to 1.12, confirming an overestimation of integrated flux, as well as the origin of this effect.
We also observe that the integrated flux is overestimated especially at small fluxes, in fact we repeated the calculation of the average ratio
between the measured and the inserted integrated flux for sources brighter than 10\,Jy and found that this average is equal to 1.05.
We have calculated the completeness and reliability also in this case. 
The completeness depends on the flux of our sources, as we have simulated sources as faint as 2\,Jy, but set a detection
threshold of 5 times the noise which has an average amplitude of 1\,Jy.
The overall completeness is 85.2 percent. If we restrict to sources brighter than 5\,Jy the completeness is equal to 98.4 percent,
for sources brighter than 6\,Jy it is 99.8 percent.
The overall reliability is 85.8 percent. Considering only bright sources, less likely to be noise peaks, it rises to 98.4 percent for
sources brighter than 5\,Jy, and to 99.8 percent for sources brighter than 6\,Jy.

\begin{table}
\centering
\begin{tabular}{c|c|c|c|c|}
\cline{2-5}
& \multicolumn{2}{|c|}{No Noise}	& \multicolumn{2}{|c|}{With Noise} \\
\cline{2-5}
& Average	& Scatter	& Average & Scatter \\
\hline
\multicolumn{1}{|c|}{Peak flux ratio}	& 0.997	& 0.009	& 1.006	& 0.259 \\
\multicolumn{1}{|c|}{Semi major axis ratio} & 0.999	& 0.004	& 1.112	& 0.547 \\
\multicolumn{1}{|c|}{Semi minor axis ratio} & 0.999	& 0.004	& 1.001	& 0.163 \\
\multicolumn{1}{|c|}{Position angle Difference} & 9\,deg	& 40\,deg	& 66\,deg	& 75\,deg \\
\hline
\end{tabular}
\caption{
Average values and scatter around them of the ratios between the measured and the input values of the peak flux,
semi major and semi minor axes, and the difference between the measured and the input values of the position angle of all sources.
In the left half we report the result of the test when sources are simulated on top of a zero background
(as in Figure~\ref{fig:valid_flux_nonoise}).
In the right half the results refer to the test when sources are simulated on top of correlated noise
(as in Figure~\ref{fig:valid_flux_noise}).
The scatters are reported at 1$\sigma$ level.
}
\label{tab:ratios}
\end{table}

As the sources simulated for these tests are all point sources and are created to have the same shape as the restoring beam,
we have also tested the use of the forced beam option, which forces all the detected sources to have the same shape
(major and minor axis and position angle of a 2-D Gaussian) of the restoring beam.
We obtain the same results when we use the forced beam option and when we perform a blind extraction, which confirms
that PySE is properly fitting the shape of extracted sources.

\begin{figure}[htbp]
\centering
\includegraphics[scale=0.5,viewport=0 0 530 410, clip]{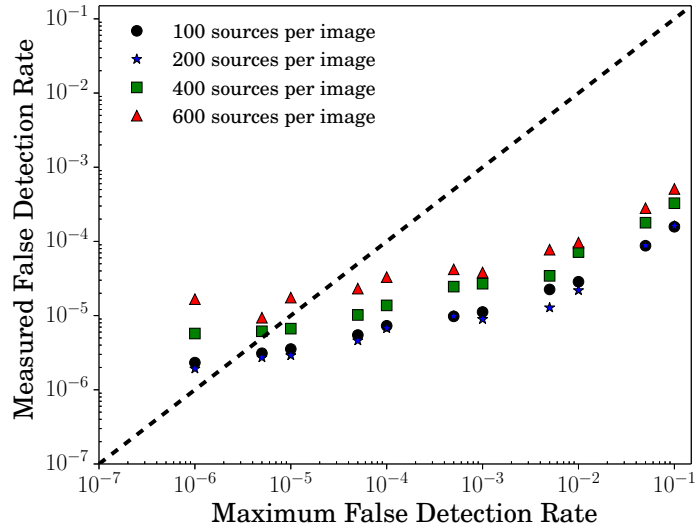}
\caption{
Results of the test of the False Detection Rate algorithm. We note that the recovered value of the False Detection Rate is
systematically smaller for smaller source density, implying that the success of the algorithm depends on the density of
sources in the image.
The dashed line represents the one-to-one relation. We can see that false detection rates as low as 10$^{-6}$ are not achieved.
}
\label{fig:fdr}
\end{figure}

In order to test the FDR algorithm in PySE we
created 10 different noise maps and for each of them we created 50 different images containing 100, 200, 400 and 600 sources
per image to test if the source density would affect the performance of the FDR.
The procedure to create the images used for this test is the same we used earlier, i.e., maps of the third type in
Section~\ref{sec:Hanno}. The source fluxes varied between 1 and 20\,Jy. We tested the performance by varying the maximum
percentage of false detections ($\alpha$), with values of 0.1, 0.05, 0.01, 0.005, 0.001, 0.0005, 0.0001, 0.00005, 0.00001,
0.000005, and 0.000001. We then compared the output source list of PySE with the input catalog and calculated the percentage
of falsely detected pixels we obtained. The results of this test are shown in Figure~\ref{fig:fdr}.
We can see that a fraction of less than 10$^{-6}$ false positives was not achieved.
Moreover, we note the measured value of the False Detection Rate is systematically smaller for smaller source density,
and with 400 and 600 sources per image, a fraction of less than 10$^{-5}$ false positives was not achieved.
This is due to the fact that a higher ratio between the number of background pixels and source pixels helps obtaining a
more accurate noise map with which the FDR algorithm calculates the threshold to be used in the source extraction.
This implies that when using the FDR algorithm one has to be aware of the density of sources in the analyzed field in order to
use an adequate limit that the algorithm can support.

The final test we did concerned the accuracy of the deblending algorithm, which separates two very close sources in
an image. It depends on the distance between sources, on the fluxes of the blended sources with respect to the detection
threshold, and on the ratio of the fluxes of two neighbouring sources. To perform this test we created maps inserting
Gaussian noise with an amplitude of 1\,Jy beam$^{-1}$ and 64 pairs of sources per map with different flux ratios and
different separations. The flux of the weakest source
was always 20\,Jy and the flux ratios were logarithmically spaced between 1 and 1000. 
This is done to avoid cases where the fainter sources might not be detected due to noise fluctuations; moreover, the ratio
between the fluxes of the two components is the most important quantity in disentangling them, together with the angular
separation. For every flux ratio 20 maps were processed, each map having a fixed separation between the two sources in
each of the 64 pairs. The separations corresponding to each of the 20 maps were linearly spaced between 2 and 7
Gaussian widths. For the source detection we chose a 10\,$\sigma$ detection threshold and a 3\,$\sigma$ analysis
threshold. The conclusion was that up to about 30 percent more sources can be detected with the deblending algorithm
than when it is not used.

For further details on these tests, see chapter 3 of \citet[][]{Spreeuw:2010}.

\subsection{PySE parameters in the Transient Pipeline}
\label{sec:pyse_params}
One of the most important reasons for developing PySE is its implementation into the LOFAR {\sc TraP}
\footnote{
Although {\sc TraP} was designed and developed within the LOFAR TKP, it has now been used with data from other telescopes,
e.g., MWA \citep{transsel_Rowlinson2016} and VLITE \citep{Polisensky2016}.}.
We have performed a study of the PySE source finder software to determine the best values of the parameters
that should be used in the {\sc TraP} version 3.0.
We have done tests on both simulated and real LOFAR maps.
The simulations used in this case were created following the method described in Section~\ref{sec:Dario}.

The images used for this test are not as simple as the ones used so far because they are possibly affected by calibration
and imaging errors.
This is to emulate the type of real image that {\sc TraP} analyses.
We have demonstrated in Section \ref{sec:validate_pyse} that the accuracy of PySE in recovering the position of the extracted
sources is very good and that the flux recovery is
affected especially by the noise RMS.
Errors beyond these limits are due to the effects previously listed that affect the accuracy of the recovered parameters.
We have used PySE and a tool to do the source matching between the output of PySE and either the sky model
from existing catalogues in the case of the real maps, or the input model in the case of the simulated maps. This
tool uses a nearest neighbour criterion for the source association,
with a cutoff at 3 source HWHM for the association.
The final choice of cutoff is rather dependent on the volume and characteristics of the data stream used. We have found that
for modest amounts of radio data (of order 10$^4$ images with each of order 10$^6$ pixels), an 8$\sigma$ limit suffices to
reduce the number of false positives enough that one can manually investigate their nature, and separate them from
real sources. However, far away from the mean the distribution of pixel values is clearly non-gaussian, and the data
artefacts vary from instrument to instrument, so we have not yet found automated ways of setting the cutoff - it needs to be
done based on close inspection of the data.
The simulation images we have analysed have a resolution of about 6~arcmin, with a pixel scale of about 1~arcmin per pixel.
This means that a point source would have a full width at half maximum of about 6 pixels.
For such configuration, we found that the best value for the grid size parameter in PySE is 50 pixels because it is larger
than all the sources in the field, and not coarse either, as explained in Section~\ref{sec:run_pyse}.
We would like to stress that the best value for the grid size might be very different than the one we quote here, depending
on the pixel scale and the beam parameters of the dataset, and the eventual presence of extended emission.

In the {\sc TraP}, the association between extracted sources and catalogue sources is performed using the De Ruiter
Radius criterion \citep[][]{deRuiter:1977}. 
The De Ruiter Radius takes into account the coordinates of the source to be associated and their errors:

\begin{equation}
\centering
R_{DR} = \left(\frac{\Delta[\alpha\,\cdot\,cos(\delta)]^{2}}{\sigma_{\alpha}^{2}} + \frac{\Delta (\delta)^{2}}{\sigma_{\delta}^{2}}\right)^{1/2} \: .
\label{eq:de_ruiter_radius}
\end{equation}

\noindent $\Delta$[$\alpha\,\cdot$\,cos($\delta$)] and $\Delta$($\delta$) represent the difference between the coordinates
of the same source as extracted by the source finder and as tabulated in the catalog, and $\sigma_{\alpha}$ and
$\sigma_{\delta}$ represent
the errors in the measurements of the coordinates of the source in the extraction by the source finder.
As the De Ruiter Radius is inversely proportional to the position errors, this means that sources of which positions are
measured with great precision will have large values for the De Ruiter Radius. This affects especially bright sources because
their positions can be measured with better accuracy than fainter ones, and can be a problem because it can happen that
very bright sources in an image are not associated.
Sources can be detected not at their nominal position because of scintillation, an effect that cause the position of sources
to move.
In the {\sc TraP} sources with a De Ruiter Radius lower than 5.68 are associated, otherwise they are not. This choice corresponds
to the probability of mis-associating sources of 1:10$^{6}$.
In our source association results we looked for sources that should be associated but have a De Ruiter Radius above 5.68.
We then added systematic uncertainties of $1/4$ pixel, $1/2$ pixel and $1$ pixel size, and examined its effect on the De
Ruiter Radius. We concluded that a systematic error of 1/2 pixel is sufficient to associate the brightest sources correctly.
In Section \ref{sec:validate_pyse} we have demonstrated PySE is very accurate in recovering the position
of the extracted, therefore the missing associations are due to other effects, such as scintillation due to the ionosphere.
More realistic simulations, including ionospheric effects, will be necessary to better determine the systematic errors needed
in our pipeline.

As part of our tests we also checked the flux recovery in both our simulated and real maps. All our simulated maps indicated
that the flux uncertainty for the sources in there was $\sim$20 percent
for all fluxes.
This amount can vary for a different dataset.
In Section \ref{sec:validate_pyse} we proved that the accuracy of PySE in the flux recovery is affected
by the amplitude of the standard deviation of the background noise, an effect that is not proportional to the flux of the source,
and by the background mean estimation.
The uncertainty measured in real maps is therefore caused by a combination of calibration errors, imaging artefacts and
ionospheric effects.

Finally, we demonstrated the functioning of the FDR algorithm, and we have proven that its the performance depends
on the source density.
We have therefore decided not to use it as a standard input in the {\sc TraP} because we do not know the number
density of sources
in the images a priori, and therefore we cannot be sure if the maximum fraction of false detections will be met.

\section{Speed Performance of PySE}
\label{sec:performance_results}
Experiments were carried out to evaluate the run-time of a single-processor execution of PySE when images of
different sizes, with different numbers of sources, are used and the speedup that can be gained on the same
images when multiple concurrent processes are used. For these tests we used only images that were simulated
using the technique described in Section~\ref{sec:Hugh}.
Sizes, number of sources and run-times of the images we used in this Section are reported in Table~\ref{tab:widths}.

PySE was run on a node of the LOFAR cluster which has 8 CPUs for parallel runs, using 
the parameters determined to be the best for standard LOFAR images (see Section~\ref{sec:pyse_params}).
In Table~\ref{tab:architecture} we report the specifications of the machine on which the performance tests were executed.

\begin{table}
\centering
\begin{tabular}{ll}
Architecture & x86$\_$64 \\
CPU op-mode(s) & 32-bit, 64-bit \\
CPU(s) & 8 \\
Thread(s) per core & 1 \\
Core(s) per socket & 4 \\
CPU socket(s) & 2 \\
NUMA node(s) & 1 \\
Vendor ID & GenuineIntel \\
CPU family & 6 \\
Model & 23 \\
Stepping & 10 \\
CPU MHz & 2003.000 \\
Virtualisation & VT-x \\
L1d cache & 32K \\
L1i cache & 32K \\
L2 cache & 6144K \\
\end{tabular}
\caption{The specification of the machine on which PySE was run for performance tests.}
\label{tab:architecture}
\end{table}

Only the run-time will be reported for all tests. There are two run-times recorded for each image: one is the total program
run-time and the other is the run-time of the parallel sections only. The PySE experiments varying the number of sources use
the former; the latter is relevant only for the speedup tests.

The sections of the algorithm that are parallelized are:
\begin{itemize}
\item Calculation of the background noise in different boxes. The analysis of different boxes can be shared among processes.
\item Extraction of islands of pixels to be analyzed. Once a peak over the detection threshold is found, neighboring pixels are
associated to the same island. This step includes deblending sources as well. The analysis of different islands can be shared
among processes.
\item Extraction of source parameters by Gaussian fitting. Once a set of true sources are found, their parameters are determined.
The analysis of different islands can be shared among processes.
\end{itemize}

We note that there is another factor that can affect the speed performance of PySE is the presence of blended sources,
i.e. more sources falling in the same island.
PySE is able to distinguish them by applying a series of logarithmically spaced sub-thresholds that will separate
the two peaks.
We did not quantify how it would affect the performance of the algorithm because it is difficult to quantify the amount
of blending.
A series of factors are playing a role here. For example, how close the peaks are (both in space and in relative brightness)
and how many peaks are blended in the same island.
The most important parameter to take into account here is the number of deb lending thresholds to be used.
The more sources are blended, the higher the number of sub-threshold should be to distinguish them, the slower the run will be.

\subsection{Results}
\subsubsection{Number of Sources}
\begin{figure}
\centering
\includegraphics[scale=0.5,viewport = 5 10 515 405, clip]{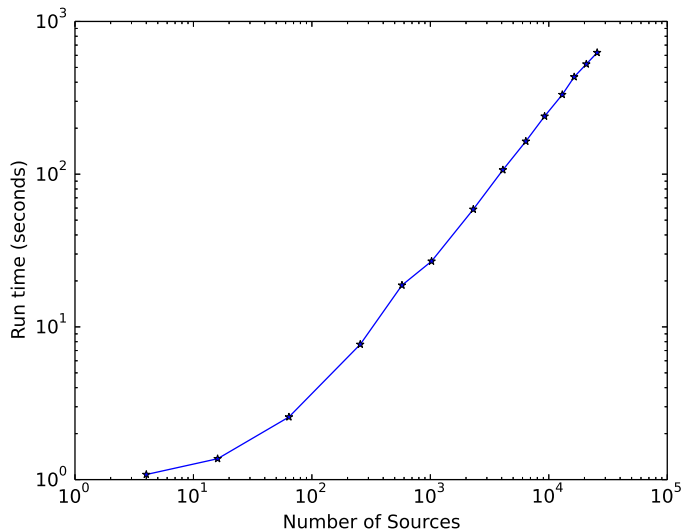}
\caption{Run-time of PySE as a function of the number of sources in the image.
The run-time increases almost linearly with the number of sources, for a number of sources greater than $\sim$100.
The slope in the log-log scale between 64 and 25600 sources is equal to 0.92.
}
\label{fig:scalings}
\end{figure}

Figure~\ref{fig:scalings} shows how the run-time of non-parallel (single process) PySE increases as the number
of sources increases. The graph shows a close-to linear relationship for a number of sources greater than $\sim$100.
This means that the overhead time is negligible in most of the datapoints.
In order to calculate the overhead time, we ran an instance of PySE on an image containing a single source
and found a runtime equal to 0.95 seconds, most of this being time for startup, image loading, background
estimation, and other initial and final tasks that are not parallelised.
A linear relationship is the best that can be expected, and PySE approximately reaches this, in fact the slope of the
curve in log-log space for a number of sources greater than 64 is equal to 0.92.

We also note that for a number of sources equal to 256 (about twice the amount we expect in a typical LOFAR image)
the run time is 7.68 seconds, larger than the requirements we set (3\,seconds), but if we extrapolate the value we expect
for 100 sources, it is close to the required speed. Further parallelism in the code can help speeding up the code further and
fulfill this requirement as well.

\subsubsection{Parallel speedup}
\begin{figure}
\centering
\includegraphics[scale=0.5,viewport = 7 10 500 390, clip]{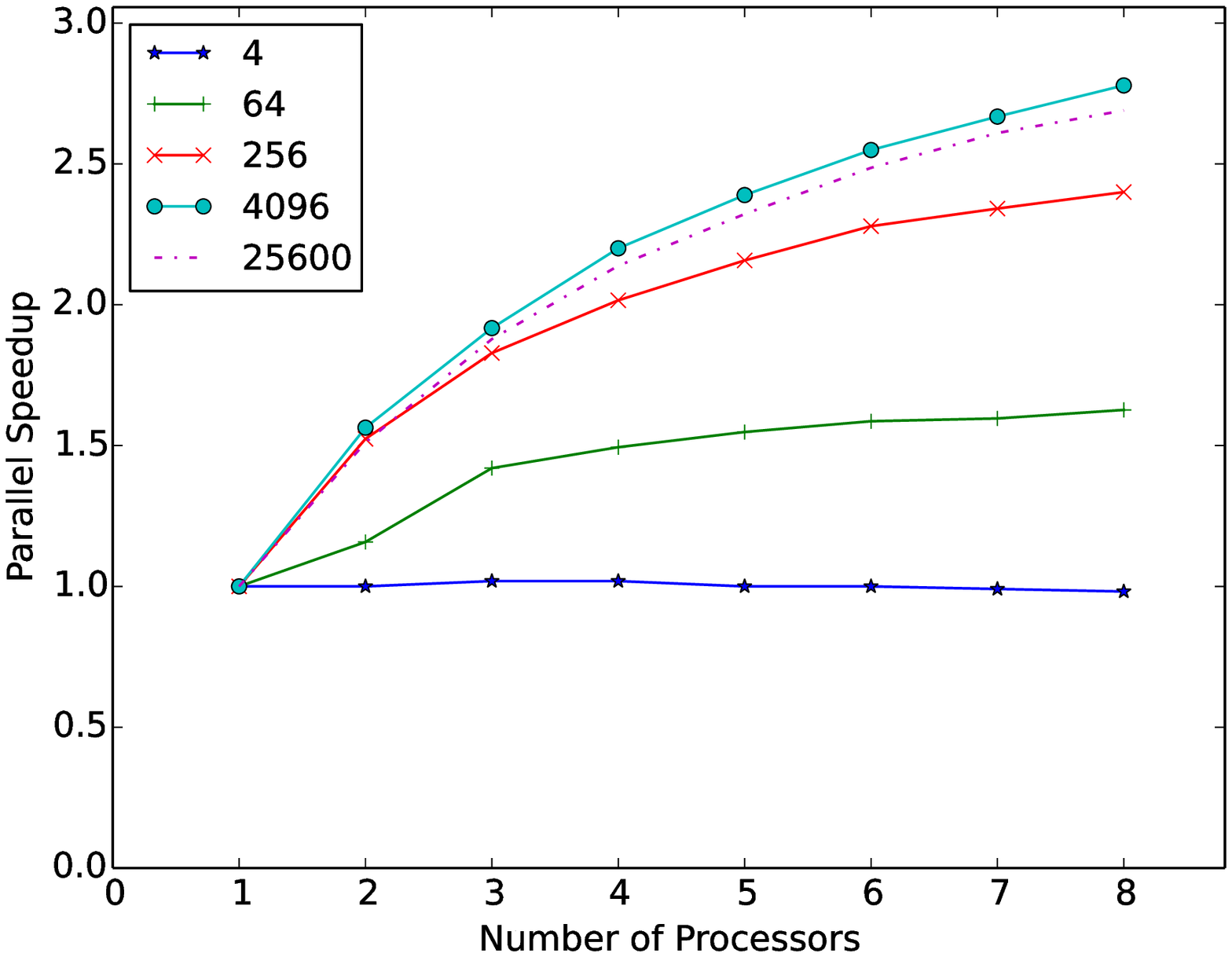}
\includegraphics[scale=0.5,viewport = 5 10 488 380, clip]{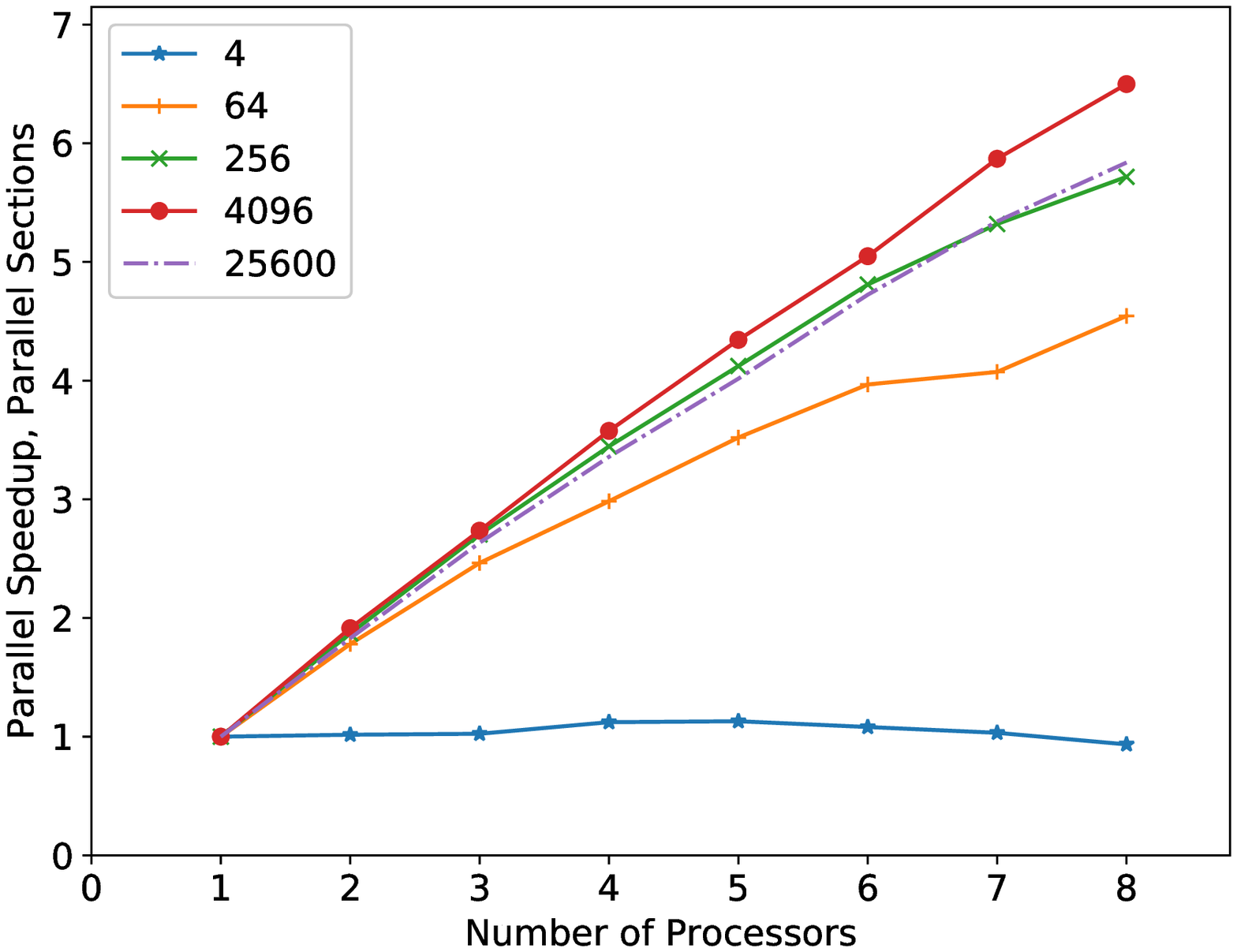}
\caption{
The top panel shows the total speed-up of PySE that can be gained when PySE is run using multiple
concurrent processes as a function of the number of sources.
The bottom panel regards shows the same relationship, but reporting the speed up of the parallel sections only.}
\label{fig:speedups}
\end{figure}

The top panel of Figure~\ref{fig:speedups} shows the total execution-time speedup that can be gained by increasing the number
of processes when PySE is run on a sample of the images used for the previous tests. The parallel speedup is
defined as the runtime for 1 processor divided by the runtime for several processes. Ideally, when N processes
are used, the speedup should be N, and the program runs N times faster. An ideal speedup is not achieved in
Figure~\ref{fig:speedups}, mainly due to the fact that only parts of the program can be parallelised.
Larger images with more sources show more speedup, since for smaller images the cost of managing
parallel processes outweighs any benefit that may be obtained, as can be seen for the image containing 4 sources,
where there is no speedup.

The bottom panel of Figure~\ref{fig:speedups} shows the speedup for just the sections of code that have been parallelised.
When there is a small number of sources, the overhead of managing the parallel code sections means that little
speedup is achieved. However, with a large number of sources, the overheads are small compared to the amount
of work to be done, and sharing the work among several processes reduces the run-time considerably.
For example, when there are 4096 sources in the image, and 8 processors over which to share the work, the run-time
is reduced by a factor of 6.3.

It is interesting to note that it is not the largest image (25600 sources) that shows the best speedup (that is at 4096 sources);
this is because the very largest images meant the operating system started paging, thus degrading performance. In fact,
an image size of $\sim$~14000$^{2}$ pixels (30000 sources), is the largest image that can be processed, depending
on system load.

\section{Comparing PySE to other source finding software}
\label{sec:discussion}
\citet{sfchallenge_Hopkins2014} conducted the ASKAP/EMU Source Finder Data Challenge comparing the accuracy of
many source finders when run on three simulated radio maps containing a mixture of blended,
unblended, point and extended sources.
Some of the source finding tools that have been tested within the Data Challenge were created to work on radio
images, whereas others were designed to work on images at other wavelengths. Despite this, all were tested on the same
simulated radio images.
The test consisted of three different maps, testing various aspects of the algorithms.
The first map (Challenge 1) contained a large number of bright sources with constant number density.
The second map (Challenge 2) showed fewer sources but with varying number density across the image in order to
mimic the clustering observed in reality. There were more faint sources  in this map, to mimic real source counts.
Both these maps contained only point sources.
The third map (Challenge 3) had the same sources as the second but 20 percent of them were turned into extended ones.
These maps were simulations of ASKAP observations.
As already mentioned, PySE was designed to detect point sources,
and for this reason we decided to focus on retrieving those also in Challenge 3, despite knowing extended emission
regions were present.
PySE was run using detection and analysis thresholds of 5$\sigma$ and 3$\sigma$ respectively.
Square cells of side 50 pixels were used for calculating the background and noise maps in Challenges 1 and 2;
30 pixel squares were used for Challenge 3.
In each case, we used the option to constrain the shape of the extracted sources to be equal to the restoring beam
and to decompose sources lying within the same island.

The results of these tests are summarised in Figures 5, 6 and 7 of \citet{sfchallenge_Hopkins2014}.
PySE achieves a completeness close to 100 percent for sources brighter than 60\,mJy in Challenge 1 and a reliability
of 100 percent for sources brighter than 40\,mJy. In this map the background noise was about 10\,mJy/beam.
In Challenge 2, where fewer sources were present, the completeness reaches 100 percent at 7\,mJy and the reliability
at 4\,mJy. In this map the background noise was about 1\,mJy/beam.
In Challenge 3 PySE did not try to fit the extended emission; despite this, it reaches a completeness of more than 80 percent,
meaning that PySE could recognise the brightest part of some of the extended sources as well. The reliability of PySE in
Challenge 3 is about 100 percent for sources brighter than 4\,mJy. Also in this case, the background noise was about
1\,mJy/beam.
PySE has been shown to be performing well both in terms of completeness and reliability of the extracted sources, with
respect to other source finder algorithms,
achieving completeness and reliability close to 100 percent when the signal-to-noise ratio is higher than about 4 in all
three challenges.

\section{Conclusions}
\label{sec:conclusions}

PySE was developed as the source finding algorithm for the LOFAR Transients Key Project in order to extract point sources
as quickly and as accurately as possible.
\citet{sfchallenge_Hopkins2014} has shown PySE to find sources with a high hit rate and low false positive rate.
We have shown it to recover sources with accurate locations, fluxes and shape.

PySE has been successfully incorporated as the standard source finding tool of the LOFAR
Transients Pipeline. It is able to perform the source extraction in real time, and it is able to cope with the throughput
of images generated by LOFAR, thanks to ability of the {\sc TraP} to run multiple instances in parallel.

PySE was designed to efficiently detect and deal with point sources within the {\sc TraP}, for example, it can force the
shape of the extracted sources to be the same as the one of the restoring beam.
Another important option of PySE is the forced extraction. This allows the user to specify a list of coordinates where
PySE must extract the flux of a point source even if a source at that location has not been found. This is very important in order
to monitor the flux of a newly detected source.

We have analysed the usage of the False Detection Rate algorithm and showed that its performance is varying
according to the source density of the field. For this reason we have decided not to use it within the {\sc TraP},
although we maintain this functionality for when PySE is operated as a standalone tool.

In real maps the errors in the positions and the fluxes recovered by PySE are much larger than the ones we measured on
simulated maps. The origin of these larger errors is attributed to factors other than PySE, such as ionospheric effects,
calibration errors and imaging artefacts.

We have demonstrated that the speed of PySE is scaling near-linearly for large numbers of sources when the overhead
time becomes negligible. Moreover, there are parts of the code that can be parallelised, increasing the performance
of the code further.

A comparison of the accuracy of PySE with other source finders has been performed
by the ASKAP Source Finder Challenge (Hopkins et al., 2015).
PySE has been shown to perform well compared to other algorithms in completeness, reliability and accuracy of the recovered
source parameters.

\section*{Ackowledgments}
LOFAR, the Low Frequency Array designed and constructed by ASTRON, has facilities in several countries, that are
owned by various parties (each with their own funding sources), and that are collectively operated by the International
LOFAR Telescope (ILT) foundation under a joint scientific policy.
DC, RAMJW, AR, GM. acknowledge support from the European Research Council Advanced Grant
247295 ``AARTFAAC''.
HG conducted this work at CEA Saclay, France, and acknowledges financial support from the UnivEarthS
Labex program of Sorbonne Paris Cit\'e (ANR-10-LABX-0023 and ANR-11-IDEX-0005-02).
This work utilises a number of {\sc Python} libraries, including the {\sc Matplotlib} plotting libraries \citep{Hunter:2007},
{\sc NumPy} \citep{Walt:2011} and {\sc SciPy} \citep{Jones:2001}.

\appendix
\section{List of PySE options}
\label{app:option}

The options that can be used when running PySE are:\newline
\texttt{-{}-fdr}: use of the False Detection Rate (FDR) algorithm to determine the
detection threshold to be used in the source extraction process.\newline
\texttt{-{}-alpha}: maximum allowed percentage of false detections in the FDR.\newline
\texttt{-{}-detection}: detection threshold, given as a multiple of the noise. The detection threshold is a factor that,
multiplied by the noise map, sets the minimum flux density that a pixel must have to be recognised as a source.\newline
\texttt{-{}-analysis}: analysis threshold, given as a multiple of the noise. Once a peak is detected, neighbouring
pixels will be analysed and if they lie above the analysis threshold they are grouped as part of the same source.\newline
\texttt{-{}-grid}: size of the squares in which the image is divided, in pixels. The image is divided into pieces in which
the background and noise are calculated and then interpolated to create the background map on top of which sources are
detected. This parameter is very important because the grid has to be not too fine, otherwise a single source can
dominate the background in certain parts, nor too coarse otherwise variations in background and noise across the
image will not be properly traced.\newline
\texttt{-{}-margin}: margin on each side of the image that will not be analysed, in pixels. \newline
\texttt{-{}-radius}: radius from the centre of the image that will be analysed, in pixels.\newline
\texttt{-{}-detection-image}: finds sources on a different image than the one where fluxes are extracted. To compensate
for the increasing noise towards the edges, it is possible to use a non-primary beam corrected image as image in
which sources are detected and then a primary beam corrected one for the calculation of their fluxes.\newline
\texttt{-{}-deblend}: use the deblending algorithm. This option requests that the program disentangle two or
more sources lying in the same island. An island is a group of pixels with fluxes above the analysis threshold,
with at least one pixel above the detection threshold.\newline
\texttt{-{}-deblend-thresholds}: number of logarithmically spaced deblending subthresholds.\newline
\texttt{-{}-force-beam}: force fit axis lengths to beam size. This option does not allow the software to fit the shape of
the sources
but instead, when a source is found, its shape is immediately fixed to the shape of the beam.
This means that using this option every source that is extracted is assumed to be a point source. \newline
\texttt{-{}-bmaj}: major axis of the beam, in degrees.\newline
\texttt{-{}-bmin}: minor axis of the beam, in degrees.\newline
\texttt{-{}-bpa}: position angle of the beam, in degrees.\newline
\texttt{-{}-fixed}: specify coordinates where a source is assumed to be, even if a source at that location has not been
detected. This is very important in case a transient source appears because we want to build its light curve from
the beginning of our observations and after its disappearance. Coordinates are given as a list in J2000 format,
in decimal degrees.\newline
\texttt{-{}-fixed-list}: specify a file containing a list of positions to force the source extraction.\newline
\texttt{-{}-ffbox-in-beampix}: specify positional freedom/error-box size for forced source extraction. Given as a multiple of beam width.\newline
\texttt{-{}-regions}: generate DS9 region file(s).\newline
\texttt{-{}-csv}: generate csv text file containing information about the extracted sources.\newline
\texttt{-{}-sky model}: generate sky model file that could be used in BBS or in Default Pre-Processing Pipeline for self calibration,
containing information about the extracted sources.\newline
\texttt{-{}-residuals}: generate residual maps to check whether the source extraction was successful.\newline
\texttt{-{}-islands}: generate island maps to check where the extracted sources are.\newline
\texttt{-{}-rmsmap}: generate RMS map of the background to check the noise level in the image.\newline
\texttt{-{}-sigmap}: generate significance map to check the signal-to-noise ratio of each pixel compared to the local RMS noise.

A review of the program and its options is given in the PySE
Manual\footnote{http://tkp.readthedocs.org/en/r3.0/tools/pyse.html}.

\section*{References}

\bibliographystyle{elsarticle-harv}
\bibliography{/Users/dariocarbone/Documents/carbone_bibliography_ac.bib}

\end{document}